\documentclass{webofc}
\usepackage[varg]{txfonts}   

\usepackage{lineno}
\usepackage{hyperref}
\hypersetup{
  colorlinks   = true, 
  urlcolor     = brown, 
  linkcolor    = black, 
  citecolor   = black 
}

\begin{document}

\title{Probing the nature of the QCD phase transition with higher-order net-proton number fluctuation and local parton density fluctuation measurements at RHIC-STAR}

\author{\firstname{Dylan} \lastname{Neff}\inst{1}\fnsep\thanks{Now at CEA Saclay-Paris, \email{dylan.neff@cea.fr}}
     for the STAR collaboration
}

\institute{University of California Los Angeles}

\abstract{
  The moments of proton and net-proton multiplicity distributions are observables expected to be sensitive to the QCD critical point and the nature of the QCD phase transition from QGP to hadron gas. Hyper-order cumulants are measured in wide centrality bins in STAR BES-I data and found to be qualitatively consistent with trends predicted by lattice QCD which finds a cross-over phase transition at low $\mu_\text{B}$. Data collected at $\sqrt{s_{NN}}=3$ GeV in BES-II exhibit trends opposite of those observed in higher energy collisions which may suggest the dominance of hadronic interactions at this energy. The variance of proton multiplicity distributions in azimuthal partitions is measured to search for signals of clustering indicative of a first-order phase transition. A strong dependence on the event multiplicity is observed. This dependence is independent of energy in AMPT while in STAR data a significant trend with energy is observed.
  




  
}

\maketitle

\section{Introduction}

A primary goal of the RHIC Beam Energy Scan program is to study the nature of the transition from Quark Gluon Plasma to hadron gas. Lattice QCD (LQCD) has established that this transition is a cross-over at vanishing baryon chemical potential ($\mu_\text{B}$)~\cite{lattice}. Model calculations have suggested that at large $\mu_\text{B}$ the transition may become first-order~\cite{first-order1, first-order2}, with a critical point marking the boundary between these two regions. Lacking first principle calculations in this high $\mu_\text{B}$ regime, we rely on experiment to search for signatures of a critical point in the QCD phase diagram.

The existence of a critical point may be inferred through deviation from cross-over behavior indicative of the onset of a first-order transition. Cumulants of net-proton multiplicity distributions, proxies for net baryon number, provide a sensitive probe of the nature of the phase transition~\cite{cumulants1, cumulants3, cumulants4}. Higher-order cumulants of these distributions can be measured and compared to trends from lattice QCD calculations valid at low $\mu_\text{B}$. Deviations from lattice expectations may indicate the end of the cross-over regime. It is also possible to search for indications of first-order behavior directly in the azimuthal correlation between protons. Coordinate space clumping is a characteristic signature of first-order phase transitions which, if present in a hypothetical QCD first-order transition, may be translated into enhanced positive correlation between the momenta of final state protons~\cite{clustering}.

We utilize Au+Au collision data from STAR Beam Energy Scan I (BES-I) along with the data set at $\sqrt{s_{NN}}=3$ GeV from the fixed-target program of BES-II to probe the nature of the QCD phase transition at large $\mu_\text{B}$.

\section{Hyper-order cumulants of proton multiplicity distributions}

\begin{figure}[ht]
\centering
\includegraphics[width=\linewidth]{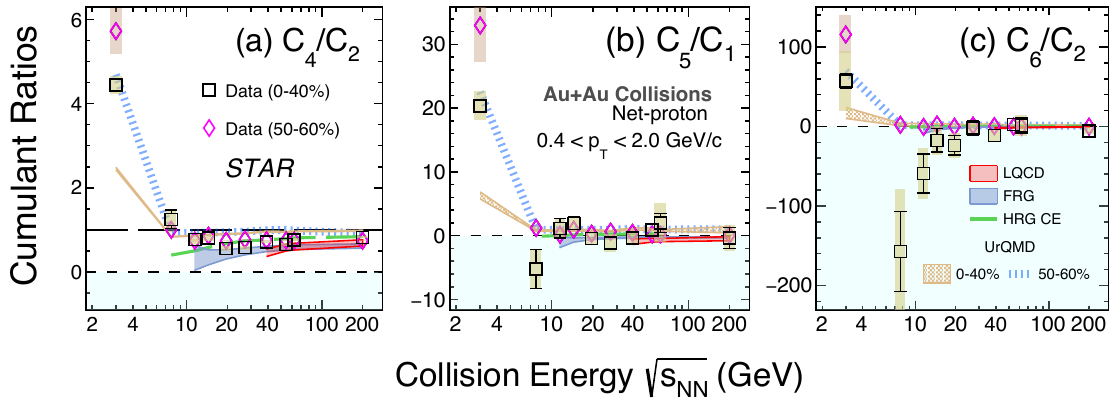}
\caption{Cumulant ratios of the net-proton distribution in BES-I data from 7.7 - 200 GeV and BES-II data at 3 GeV. $C_4/C_2$ (a), $C_5/C_1$ (b), and $C_6/C_2$ (c) are measured in the 0-40\% most central events (squares) and compared with the 50-60\% most central events (diamonds). LQCD~\cite{lqcd}, FRG~\cite{frg}, UrQMD~\cite{urqmd}, and HRG CE~\cite{hrg} model calculations are shown for comparison.}
\label{fig:cumulant_ratios}
\end{figure}

\noindent Higher-order ($C_3$ and $C_4$) and hyper-order ($C_5$ and $C_6$) cumulants of net-proton distributions are measured in BES-I data and their ratios are plotted in Figure~\ref{fig:cumulant_ratios} as a function of center of mass collision energy. $C_4/C_2$ is positive for all energies and shows no significant deviation from the model calculations for these centrality ranges. LQCD and FRG calculations predict a negative $C_5/C_1$ and $C_6/C_2$ which become more negative with decreasing beam energy~\cite{lqcd, frg}. While no significant beam energy dependence is observed in $C_5/C_1$, $C_6/C_2$ is observed to be increasingly negative with decreasing energy between $\sqrt{s_{NN}}=7.7\text{ GeV} - 200\text{ GeV}$ --- qualitatively consistent with the trend found in lattice calculations. The 3 GeV fixed target data point is observed to be positive for all three ratios and consistent with UrQMD for $C_6/C_2$.

\begin{figure}[ht]
\centering
\includegraphics[width=\linewidth]{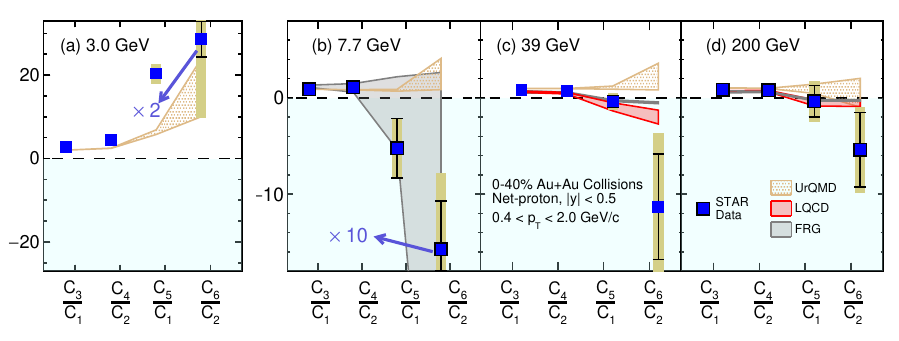}
\caption{Cumulant ratios are plotted with increasing order on the x-axis for 3 GeV (a), 7.7 GeV (b), 39 GeV (c), and 200 GeV (d) for the 0-40\% most central events. These measurements are compared with LQCD~\cite{lqcd}, FRG~\cite{frg}, and UrQMD~\cite{urqmd} model calculations.}
\label{fig:cumulant_ratio_ordering}
\end{figure}

Lattice calculations, reliable down to 39 GeV, also predict an ordering of the higher and hyper-order cumulant ratios for each energy, with the ratio decreasing as the order increases. This trend is plotted in Figure~\ref{fig:cumulant_ratio_ordering} where we find that STAR data from 7.7 to 200 GeV appear consistent with the predicted hierarchy within statistical uncertainty. The fixed target 3 GeV data seem to exhibit the opposite trend which is reproduced by UrQMD, suggesting that hadronic interactions may be dominant at this energy.

\section{Measurement of proton correlation within azimuthal partitions}


Excess clustering in the azimuthal distribution of protons may be indicative of coordinate space clumping characteristic of first-order phase transitions. To search for signals of clustering, the azimuth of each event is partitioned and the number of proton tracks within each azimuthal partition are counted. For an event with $N$ total protons in the full azimuthal acceptance, randomly distributed tracks should produce a binomial distribution in the partitioned multiplicity of fixed azimuthal width $w$ corresponding to $N$ trials and probability of success $p=w/2\pi$. 

The variance of the azimuthal multiplicity distributions compared to the uncorrelated binomial variance, $\sigma^2_{\text{binomial}}=Np(1-p)$, is sensitive to the correlation among protons. Variance larger than $\sigma^2_{\text{binomial}}$ indicates a positive correlation between protons -- a clustering signal. Variance smaller than $\sigma^2_{\text{binomial}}$ indicates negative correlation -- a repulsive interaction. We construct an observable to quantify and properly normalize the deviation of the measured variance from the binomial variance:

\begin{equation}
    \Delta \sigma^2\left(N\right) = \frac{\sigma^2(N) - \sigma^2_{\text{binomial}}(N)}{N (N - 1)}
    \label{eq:dsig2}
\end{equation}

We find that the $N(N-1)$ normalization in Equation~\ref{eq:dsig2} effectively removes $N$ dependence from $\Delta \sigma^2$ measured in STAR and model data. This justifies taking an average over $N$ which we denote as $\langle \Delta \sigma^2 \rangle$. 

\begin{figure}[hb]
\centering
\includegraphics[width=\linewidth]{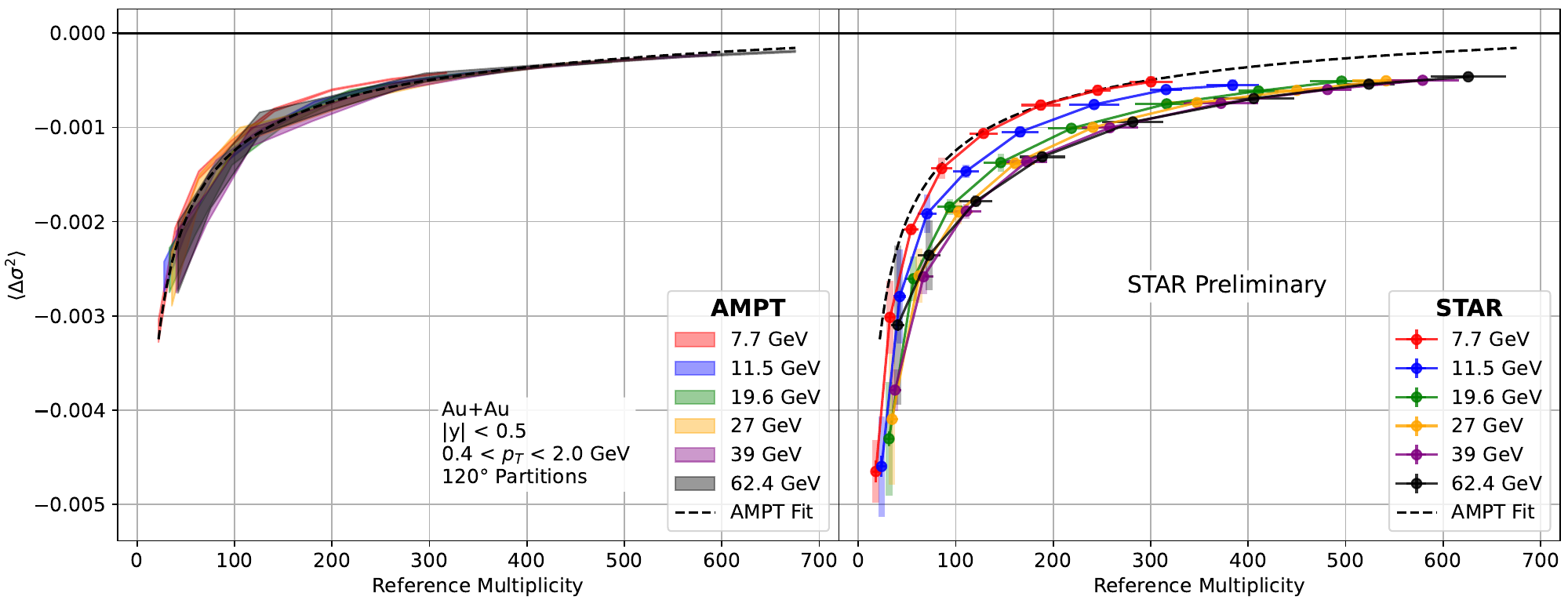}
\caption{$\langle \Delta \sigma^2 \rangle$ is plotted for each centrality class as a function of the average reference multiplicity within that centrality class for AMPT~\cite{ampt} model data on the left and STAR data on the right. A dashed line outlining the trend found in AMPT is shown in both panels to aid in comparison.}
\label{fig:dsig2_vs_refmult}
\end{figure}

\newpage
In Figure~\ref{fig:dsig2_vs_refmult} we find that $\langle \Delta \sigma^2 \rangle$ is significantly negative for all energies and centralities in both STAR and AMPT~\cite{ampt} data, indicating a repulsive interaction between protons. In addition, strong reference multiplicity dependence is observed, with the magnitude of repulsion dramatically increasing as the event multiplicity decreases. We postulate that this dependence is due to global momentum conservation, which serves as a background and obscures any possible clustering signal. We note that while for AMPT $\langle \Delta \sigma^2 \rangle$ is energy independent and all data falls on a universal curve, STAR data exhibits significant energy dependence. Higher energy data sets in STAR data appear to approach a universal curve as in AMPT, but as energy decreases $\langle \Delta \sigma^2 \rangle$ appears to become less negative. This could be consistent with a clustering signal whose magnitude increases with decreasing energy superimposed on a large energy independent background.

\section{Summary}

STAR BES-I data was utilized to probe the nature of the QCD phase transition at finite $\mu_\text{B}$. Measurement of hyper-order cumulants of the net-proton distribution between 7.7 and 200 GeV produced results qualitatively consistent with lattice QCD predictions. The hyper-order cumulant measurements in 3 GeV fixed target data deviate from trends found at higher energies and showed consistency with the UrQMD model, suggesting that hadronic interactions may be dominant at 3 GeV. A strong repulsive interaction was observed from the measurement of proton multiplicities in azimuthal partitions. This repulsion exhibited strong dependence on the event multiplicity, suggesting it may be due to momentum conservation. Energy dependence of $\langle \Delta \sigma^2 \rangle$ was observed in STAR data while being absent in AMPT.

%
%
%

\end{document}